\documentclass[12pt]{iopart}

\usepackage{graphicx}
\usepackage{iopams}
\usepackage[utf8]{inputenc}

\newcommand{\unit}[1]{\ensuremath{\, \mathrm{#1}}}

\begin{document}

\title{Crossover from photon to exciton-polariton lasing}

\author{E~Kammann$^1$, H~Ohadi$^1$, M~Maragkou$^1$, A~V~Kavokin$^1$ and P~G~Lagoudakis$^1$}%
\address{$^1$ School of Physics and Astronomy, University of Southampton, Southampton, SO17
 1BJ, United Kingdom}%

\begin{abstract} We report on a real-time observation of the crossover between
    photon and exciton-polariton lasing in a semiconductor microcavity.  Both
    lasing phases are observed at different times after a high-power excitation
    pulse. Energy-, time- and angle-resolved measurements allow for the
    transient characterization of carrier distribution and effective
    temperature. We find signatures of Bose-Einstein condensation, namely
    macroscoping occupation of the ground state and narrowing of the linewidth
    in both lasing regimes. The Bernard-Douraffourgh condition for inversion was tested and the polariton laser as well as the photon laser under continuous wave excitation were found to operate below the theoretically predicted inversion threshold.
\end{abstract}

\maketitle

Bose-Einstein condensation (BEC) of exciton-polaritons in semiconductor
microcavities
\cite{kasprzak_bose-einstein_2006,kasprzak_formation_2008,deng_condensation_2002}
and photons in dye-filled microcavities operating in the weak exciton-photon
coupling regime \cite{klaers_bose-einstein_2010} have been reported in recent
years. Unlike atomic condensates in harmonic
traps~\cite{davis_bose-einstein_1995,anderson_observation_1995}, where a
coherent state is achieved by cooling down of the bosonic thermal distribution,
the condensates (coherent states) of polaritons and photons can be formed
without thermal equilibrium for example by parametric
amplification~\cite{savvidis_angle-resonant_2000} or under nonresonant
excitation at negative detunings~\cite{kasprzak_formation_2008}. Moreover, the
spectra of vertical cavity surface emitting semiconductor lasers (VCSELs)
frequently show thermal tails coexisting with the lasing
mode~\cite{van_exter_effect_1995} that suggest thermal equilibrium of
photons~\cite{klaers_thermalization_2010}. Essentially the experimental
observations of polariton BEC in the strong coupling
regime~\cite{kasprzak_bose-einstein_2006}, to photon BEC in the weak coupling
regime~\cite{klaers_bose-einstein_2010} and weak coupling lasing or VCSEL
operation~\cite{van_exter_effect_1995,bajoni_photon_2007} have very similar
signatures.  Carriers are distributed according to the Bose-Einstein
distribution, the emission narrows in energy and the first-order spatial
coherence builds up.  Recently we showed that spontaneous symmetry breaking,
which is the Landau criterion for the phase transition can also be observed in
polariton and photon lasers~\cite{ohadi_spontaneous_2012}.

However, the physical processes by which condensation and conventional lasing
occur are fundamentally different. Condensation is a purely thermodynamic phase
transition during which the total free energy of the system is minimized,
whereas conventional lasing is a balance between the gain from inversion and the
loss in the system. In a conventional semiconductor laser, lasing occurs by the
stimulated emission of the cavity photons from the e–h plasma. Above a threshold
density, the stimulated emission becomes faster than the thermalization rate. As
the result, a dip is formed in the carrier distribution, which is called kinetic
hole burning~\cite{yamaguchi_becbcs-laser_2012}. This is because the
thermalization process can no longer supply the lost carriers at sufficient
speed. In condensation, however, the system remains thermalized while lasing.
The question whether the term BEC or lasing is appropriate for degenerate
condensates of exciton polaritons and photons is still a subject of great debate
in the scientific community~\cite{butov_behaviour_2012,snoke_polariton_2012}. 

The crossover from strong to weak coupling according to the coupled oscillator
model takes place when the exciton-photon coupling strength ($g_0$) equals half
of the difference between the decay rates of cavity photons ($\gamma_{cav}$) and
excitons ($\gamma_{exc}$)~\cite{savona_quantum_????}. This may be achieved by
changing the optical pumping strength. The exciton linewidth increases and the
oscillator strength decreases with the increase of pumping intensity, which
brings the system from strong to weak coupling. This transition is not to be
confused with the Mott transition from an exciton gas to an electron hole
plasma~\cite{kappei_direct_2005,stern_mott_2008,koch_exciton_2003,amo_interplay_2006}.
Whilst the distinction between strong and weak coupling in a microcavity is
straightforward, as the dispersion relations exhibit specific differences, it is
very hard to identify the exact point of the Mott transition by standard
spectroscopic means.  A transition to the weak-coupling regime with increasing
pumping strength in steady state has been observed by several groups and the
carrier densities at the onset of photon lasing compare well with the Mott
density~\cite{deng_polariton_2003,nelsen_lasing_2009,balili_role_2009,tsotsis_lasing_2012,
tempel_characterization_2012}. This paper completes this series as it
investigates the dynamical transition from the weak to the strong exciton-light
coupling regime in a planar semiconductor microcavity excited by a short
high-power excitation pulse. We particularly investigate the distributions of
carriers during this crossover and discuss the possibility of a BEC of photons.
We observe clear features of polariton and photon lasing and find quasi-thermal
distributions of quasiparticles in the weak- and in the strong-coupling regime,
which could imply BEC of photons and polaritons. A closer look at the temporal
dynamics and the change of the effective temperatures during the transition
provide insight into the nature of the observed lasing modes and the
thermodynamic state of the system. We further investigate the build-up of photon
lasing under continuous wave excitation and investigate the question whether the
system is inverted by means of the Bernard-Douraffourgh condition for lasing.

The system under study is a GaAs microcavity grown by molecular beam epitaxy.
Previous works have shown that lasing in the weak coupling can be observed in
this sample under continuous wave (CW) excitation~\cite{bajoni_photon_2007}, whilst
nonlinearities in the strong-coupling regime are accessible under pulsed
excitation, due to the sample overheating~\cite{maragkou_longitudinal_2010}. At
higher excitation power the emission switches to the weak-coupling regime,
similar to the observations in reference~\cite{tempel_characterization_2012}. We
show that the photon and polariton lasing occur at different times after the
excitation pulse and that above threshold photon lasing is followed by polariton
lasing. Experiments were carried out using a liquid helium cooled wide-field view
cold finger cryostat.  The exciton-cavity mode detuning was set to $-0.5
\unit{meV}$. Transform limited pulses from a femtosecond Ti:Sapph oscillator
tuned to a reflection minimum of the Bragg mirror were focused to a $30\unit{\mu
m}$ spot through an objective with a high numerical aperture ($\mathrm{NA}=0.7$). The
dispersion relation was imaged through the same objective onto the slit of a
monochromator equipped with a water cooled CCD and a streak camera with ps
resolution. For temporal resolution the momentum space was scanned across the
crossed slits of the streak camera and the monochromator.  

\begin{figure}
\centering
\includegraphics[width=\textwidth]{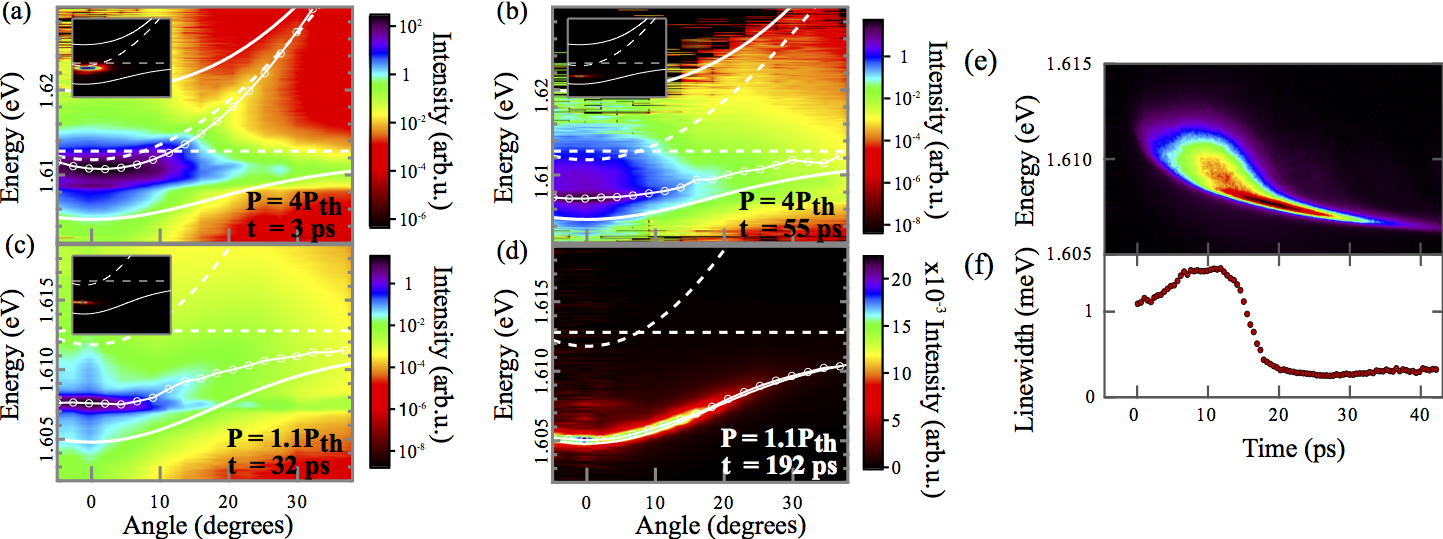}%
\caption{Dispersion images at different times and excitation powers. (a), (b),
(c), (d) Dispersions at excitation powers of $P = 4 P_{th}$ (a, b) and  $P = 1.1
P_{th}$ (c, d) at $3\unit{ps}$ (a), $55\unit{ps}$ (b), $32\unit{ps}$ (c) and
$192\unit{ps}$ (d). The insets in (a), (b), (c) show the same dispersions but in
linear color scale. At $P = 4 P_{th}$ a transition from the weak to the
strong-coupling regime is clearly observed between (a) and (b), whilst at lower
excitation powers a blue shifted exciton-polariton laser at early times relaxes
towards the modes of the linear strong-coupling regime. Images are bilinearly
interpolated.  The solid lines indicate the shape of the modes at low excitation
powers and the dashed lines are the corresponding bare exciton and cavity modes.
The white circles follow the location of maxima corresponding to the measured
photon and exciton-polariton modes. (e) Typical energy resolved evolution of the
groundstate emission intensity for $P=8 P_{th}$. (f) Linewidth as a function of time.}
\label{fig:disp}
\end{figure}

\Fref{fig:disp}a shows a snapshot of a bi-linearly interpolated image of the
microcavity dispersion $3\unit{ps}$ after optical excitation at the excitation density
$P = 4 P_{th}$ (where $P_{th}=7\unit{mW}$ is the power threshold for lasing).
The photoluminescence intensity is displayed in false-color logarithmic scale.
The inset of \fref{fig:disp}a shows the same microcavity dispersion in
false-color linear scale. Solid white lines indicate the exciton-polariton
branches in the linear regime and the dashed lines show the bare cavity and
exciton modes. Observation of the bare cavity photon dispersion confirms that
the excitation pulse brings the microcavity to the weak-coupling regime.
\Fref{fig:disp}b shows the same as \fref{fig:disp}a but at $55\unit{ps}$ after
optical excitation. White circles indicate the intensity maxima of the recorded
spectrum at each detection angle, following the cavity (a) and the polariton
mode (b). The lower exciton-polariton dispersion is uniformly blue-shifted due
to the repulsive interaction with the exciton reservoir. It is instructive to
compare these results with the lower excitation power sufficient to excite a
polariton condensate.  \Fref{fig:disp}c shows a snapshot of the microcavity
dispersion $32\unit{ps}$ after optical excitation at the excitation density $P =
1.1 P_{th}$. Similarly to \fref{fig:disp}b, a near uniformly blue-shifted lower
exciton-polariton dispersion is observed. \Fref{fig:disp}d shows a snapshot of
the microcavity dispersion at $192\unit{ps}$ under the same optical excitation,
when the exciton-polariton dispersion in the linear regime is fully recovered as
a result of the depletion of the exciton reservoir. Therefore, using
time-resolved dispersion imaging we observe the dynamics of the transition of
the microcavity eigenstates through three distinctively different regimes: from
the weak-coupling regime where we observe a bare cavity mode, to the non-linear
strong-coupling regime featuring a blue-shifted lower exciton-polariton branch,
through to the linear strong-coupling regime where the exciton-polariton
dispersion is not altered spectrally. The temporal evolution of the groundstate
energy is depicted in \fref{fig:disp}e, showing the redshift of the emission
with time.  This reflects the transition from the weak to the strong-coupling
regime and can be understood as an effect of the depletion of the carrier
reservoir. The emission linewidth reflects the coherence properties in the three
regimes. Starting from a linewidth of $\sim 1 \unit{meV}$ in the photon lasing
regime the linewidth initially increases when the system enters the transitory
regime and then narrows down to the linewidth of the polariton
laser~\cite{del_valle_dynamics_2009} (\fref{fig:disp}f).  The time-resolved
spectra and linewidth evolutions in the linear and nonlinear strong coupling
regime are given in the supplemental material 1. The time axis was rescaled to
account for the temporal distortion caused by the use of the grating (for further
information see the supplemental material 2).

\begin{figure*}
\centering
\includegraphics[width=\textwidth]{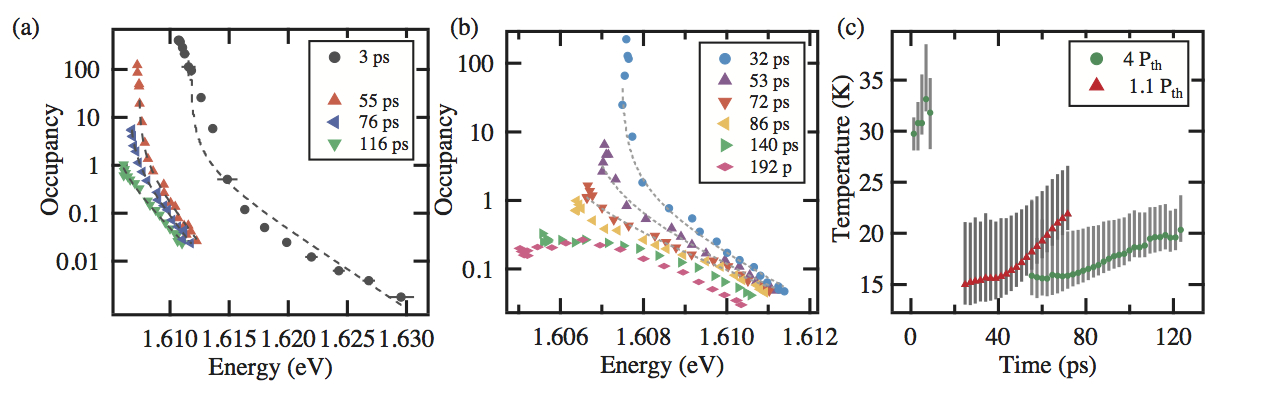}%
\caption{Energy distribution and Temperatures. (a) the occupation measured at
different energies are  given for several times after the optical
excitation for $P = 4 P_{th}$. At $3 \unit{ps}$ a photon laser at the
energy of $1.610 \unit{eV}$ coexists with thermalized photons at
$\sim30\unit{K}$ populating the cavity dispersion.  At $55 \unit{ps}$ and $76
\unit{ps}$ a Bose-Einstein distribution of exciton-polaritons with a degenerate
ground state is observed. After $116 \unit{ps}$ the occupancy of the ground
state reaches one. (b) Analysis of the spectra for $P = 1.1 P_{th}$ at different
angles and times show the coexistence of an exciton-polariton lasing mode at the ground
state with a thermalized population at early times. We map the depletion of the
condensate until the linear exciton-polariton regime,  where the distribution
exhibits a bottleneck. (c) The temporal evolution of temperature for $P = 1.1
P_{th}$ (red triangles) and $P = 4 P_{th}$ (green circles). The dashed lines in
(a) and (b) are BE fits with a spectral range of $>3k_BT$.} \label{fig:temps}
\end{figure*}

\Fref{fig:temps}a shows the occupancy as a function of energy at
different times for $P = 4 P_{th}$. At early times ($3\unit{ps}$), whilst still
in the weak-coupling regime, we observe a massively-occupied cavity mode ground
state on top of a thermalized tail of excited states. In the transitory
regime ($10-54\unit{ps}$) the dispersion cannot be mapped because the linewidth
at higher angles is strongly broadened and therefore a distribution of population
is unattainable. At later times, $55\unit{ps}$ and $76\unit{ps}$ after optical
excitation whilst in the strong-coupling regime, we observe a largely occupied
exciton-polariton ground state coexisting with a thermalized exciton-polariton
gas. After $\sim116\unit{ps}$ the ground state is no longer degenerate and the
particle distribution is close to a Boltzmann distribution.  At even later
times, the occupation of the ground state cannot be resolved as it is four
orders of magnitude lower than at the peak emission intensity.
\Fref{fig:temps}b shows successive snapshots of the emission intensity
of exciton-polaritons as a function of energy for low excitation powers ($P =
1.1 P_{th}$). We observe a largely occupied ground exciton-polariton state on
top of a thermalized exciton-polariton gas, in low-excitation nonlinear regime.
The depletion of the ground state and the loss of thermalization occur around
the same time ($\sim86\unit{ps}$) as a bottleneck builds out ($140\unit{ps}$ and
$192\unit{ps}$).  \Fref{fig:temps}c shows the temporal evolution of
temperature in the transition from photon to exciton-polariton condensate (green
circles), and from polariton condensate to a thermalized exciton-polariton gas
(red triangles). Effective temperatures were extracted by fitting a
Bose-Einstein distribution to the measured angular distribution of the emission
intensity~\cite{deng_quantum_2006} (dashed grey lines in \fref{fig:temps}a
and (b)). This analysis provides insight into the thermodynamics of the system
and how far how far away from thermal equilibrium the quasiparticles are.  The
effective temperature in the weak-coupling regime ($\sim 32 \unit{K}$) is higher
than in the strong-coupling regime ($\sim 16\unit{K}$), while in both cases the
quasiparticles remain warmer than the lattice temperature ($\sim 6\unit{K}$).
The lower effective temperature of the polariton gas reflects the longer
timescale on which they thermalize with respect to photons. At the formation
stage of the photon laser, the effective photon temperature is higher
($\sim32\unit{K}$) than the subsequent exciton-polariton gas ($\sim16\unit{K}$). 

The photon gas thermalizes via absorption and re-emission processes in the
intracavity quantum wells, similar to the mechanism of photon thermalization in
a dye-filled microcavity~\cite{klaers_thermalization_2010}. This thermalization
mechanism is analogous to the exciton-polariton thermalization in the
strong-coupling regime if the system is below the Mott transition and the
excitons are still present in the weak-coupling regime. In this case each photon
state has a finite exciton fraction even in the weak-coupling regime, which
allows for efficient interaction with phonons and other dressed photons. On the
other hand, the observed photon lasing mode occurs at much shorter times than
the usual exciton formation rates of tens of
picoseconds~\cite{kumar_picosecond_1996} in GaAs. In this case the thermalized
distribution originates from the ultrafast self-thermalization of an electron
hole plasma (tens of femtoseconds)~\cite{lin_femtosecond_1987}. Thermalization
and BEC in an ionized plasma is in principle
possible~\cite{zeldovich_bose_1969}, through Compton scattering.

Upon the formation of excitons the effective temperature approaches the lattice
temperature through carrier phonon scattering in the picosecond timescale.
Exciton-polaritons have larger exciton fraction than the photons, which is why
they interact stronger with acoustic phonons and between themselves. The cooling
of exciton-polaritons occurs on a longer timescale providing a different
temperature for the exciton-polariton gas with respect to that of the exciton
reservoir and the host lattice.  In the intermediate regime of the temporal
transition from photon to exciton-polariton laser, the distinction between the
weak and strong-coupling regime in momentum space vanishes and the energy
appears more and more red shifted, indicating broadband emission similar to the
kind observed in reference~\cite{guillet_laser_2011} or the coexistence of
polariton and photon lasing~\cite{lagoudakis_coexistence_2004}.

\begin{figure} \centering \includegraphics[width=0.9 \textwidth]{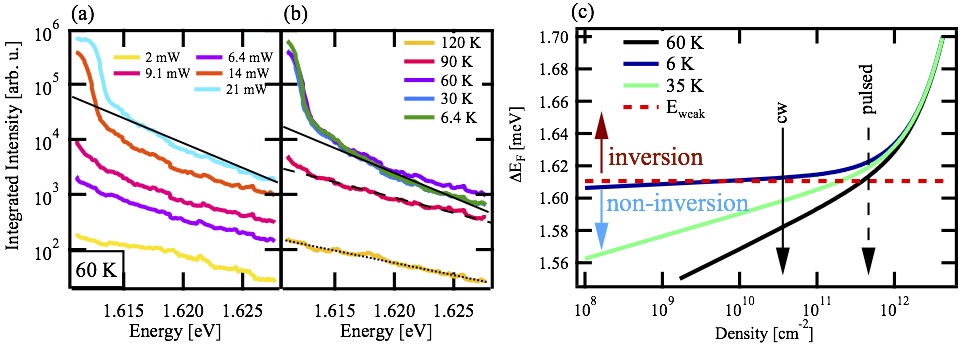}
    \caption{Photon lasing in the steady state.  (a) At $60\unit{K}$ the
        distribution of photons thermalizes at a temperature close to the
        lattice temperature and the ground state becomes macroscopically
        occupied at a certain excitation power. (b) Photon distributions for
        different cryostat temperatures. The transition can equally be induced
        by lowering of the temperature. Photon distributions follow the lattice
        temperature and the ground state becomes degenerate around $60\unit{K}$.
        While the lattice temperature can be varied down to $\sim 6\unit{K}$ the
        photon distribution does not go below $60\unit{K}$. Straight lines
        indicate Boltzmann distributions for $60\unit{K}$ (solid), $95\unit{K}$
        (dashed) and $120\unit{K}$ (dotted). (c) Difference between the Fermi
        levels of electrons and holes as a function of carrier density. For
        three different temperatures, corresponding to the cryostat temperature
    ($6\unit{K}$), the temperature in the weak coupling measured for pulsed
excitation ($35\unit{K}$) and under CW excitation ($60\unit{K}$). The arrows
indicate the experimental conditions. } \label{fig:TI} \end{figure}

The buildup of the photon laser is below the temporal resolution of our
detection apparatus. Therefore we show time-integrated photoluminescence spectra
under CW excitation at different temperatures and excitation densities.
\Fref{fig:TI}a shows pump power dependent data at $60\unit{K}$. The distribution
thermalizes at a temperature close to the lattice temperature (solid line) and
upon saturation the ground state becomes macroscopically occupied, as observed
in reference~\cite{klaers_bose-einstein_2010}.  Next we induce the transition
from a thermalized distribution to a photon laser by lowering the temperature,
to demonstrate further similarities to
atom~\cite{davis_bose-einstein_1995,anderson_observation_1995} and polariton
condensates~\cite{kasprzak_formation_2008} (\fref{fig:TI}b).  We use CW
excitation at a constant excitation power and study the emission pattern of the
cavity mode as a function of temperature. At a critical temperature of about
$90\unit{K}$ the thermal distribution achieves the degeneracy threshold and
photons start condensing at the cavity ground state.  Although such behavior is
characteristic of a thermodynamic phase transition, it is more likely due to the
change of the cavity mode energy with respect to the electron hole transition
energy in a similar fashion to VCSELs~\cite{zou_ultralow-threshold_2000}.  The
temperature of the photon gas follows the lattice temperature between
$60\unit{K}$ and $120\unit{K}$, but does not go below $60\unit{K}$ which is
slightly above the exciton binding energy in bulk
GaAs~\cite{amo_interplay_2006}. This might be an indication that the
electron-hole pairs are unbound in this case.  In the time-resolved experiments
we detect a lower temperature for the photon condensate (\fref{fig:temps}c), due
to lesser heating of the sample under pulsed excitation. 

Next we test the Bernard-Douraffourg condition for lasing~\cite
{bernard_laser_1961} by comparing the emission energy in the weak coupling
regime ($E_{weak}$) with the difference between the Fermi energies ($\Delta
E_F$) of electrons in the conduction band and holes in the valence band for the
weak-coupling regime. Conventional lasing occurs when $\Delta E_F>E_{weak}$.
Calculations of the Fermi-energies and carrier densities are provided in the
supplemental material 3. \Fref{fig:TI}c shows $\Delta E_F$ as a function of the
density of electron hole pairs. Arrows indicate the experimental conditions. The
system is close to the theoretically estimated inversion-threshold, at the onset
of photon lasing under pulsed excitation.  However, the density in the CW case
remains an order of magnitude below the inversion density predicted for
$60\unit{K}$, which is the lowest photon temperature measured under CW
excitation suggesting that photons are coalescing to a condensed state.

In conclusion, we have studied the dynamic transition from a photon to a
polariton laser after a high-power excitation pulse. Dispersion images clearly
show the transition from the weak to the strong coupling. Both regimes exhibit
the same signatures of Bose-Einstein condensation: A macroscopic occupation of
the ground state on top of a thermalized tail, narrowing of the linewidth and
narrowing of the distribution in momentum space. We have shown that the
transition to the photon laser can be induced by decreasing the temperature. The
carrier densities remain below the inversion threshold in the CW excitation
regime as well as in the strong coupling regime. The effective temperatures show
how far from thermal equilibrium the system is at different times after the
excitation pulse and the evolution of the linewidth maps the transition between
two coherent states by a passage through an incoherent state. The results
presented here as well as the recently reported observation of spontaneous
symmetry breaking and long-range order~\cite{ohadi_spontaneous_2012} calls for
further studies of condensation in the weak-coupling regime.

\ack
The authors thank Jacqueline Bloch and Aristide Lema\^{i}tre for provision of
the sample. We would like to acknowledge the FP7 ITN-Spinoptronics, ITN-Clermont
4, Royal Society and EPSRC through contract EP/F026455/1 for funding. E.K. and
H.O. acknowledge Martin Weitz for discussions. 

\section*{References}

\bibliographystyle{unsrt}
\bibliography{Refs_Kammann}
\end{document}